\documentclass[
 ,twocolumn
 ,secnumarabic
,tightenlines ,amssymb, amsmath,nobibnotes, aps, prl ]{revtex4}
\usepackage{graphicx}
\usepackage{dcolumn}
\usepackage{bm}
\usepackage{epsfig}
\begin{document}
\preprint{test1}
\title{ Palatini $f(R)$ cosmology and Noether symmetry}
\author{\textbf{Mahmood Roshan}}
\author{\textbf{Fatimah Shojai}\\ Department of Physics, University of Tehran, Tehran, Iran.}
\begin{abstract}
We study Palatini $f(R)$ cosmology using \textit{Noether symmetry
approach} for the matter dominated universe. In order to
construct a point-like Lagrangian in the flat $FRW$ space time,
we use the dynamical equivalence between $f(R)$ gravity and
scalar-tensor theories. The existence of Noether symmetry of the
cosmological $f(R)$ Lagrangian helps us to find out the form of
$f(R)$ and the exact solutions for cosmic scale factor. We show
that this symmetry always exist for $f(R)\sim R^{n}$ and the
Noether constant is a function of  the Newton's gravitational
constant and the current matter content of the universe.
 \end{abstract}
\maketitle
\section{1.\;\;\;Introduction}
The accelerated expansion  of the universe which has revealed
from the observations of  type Ia supernova \cite{1}, is one of
the biggest problems that theoretical physics faces nowadays. The
equations of motion of $GR$ can not explain such an acceleration
by standard sources of matter and energy and the mechanism that
drives this observed expansion is yet unclear. In an attempt to
justify this puzzle, some certain modifications of classical
cosmology have been proposed. It should be noted that before the
this  some modifications had been applied to standard $GR$
too. For example the scalar-tensor theory of gravity had been
proposed for improving  $GR$ from the Mach principle's point of
view \cite{2}. However, the situation is different now and this
cosmic speed up may be a serious signal of a failure of $GR$. Some
authors fallow the idea that dark energy sources such as
cosmological constant or quintessence could be responsible for
this acceleration \cite{3,4,5,6,7,8}. On the other hand, there
exist some ideas that propose modification of the Hilbert-Einstein
Lagrangian arguing that there could be nonlinear terms of Ricci
scalar in the action. The theories constructed in this way are
known as $f(R)$ gravities. These theories propose a geometric
origin for cosmic speed up without any need for sources of
dark energy. For a given $f(R)$ Lagrangian , as the $GR$ case ,
there exist two ways for constructing the gravitational theory,
namely , metric and Palatini formalisms. But unlike the $GR$ case,
this two formalisms are not equivalent, the former leads to a
system of fourth-order partial differential equations for the
metric whereas the later leads to second-order equations. It has
been shown that many different $f(R)$ Lagrangians lead to correct
cosmic acceleration \cite{9,10,11}, on the other hand, there exist
some argumentations against Palatini $f(R)$
gravity \cite{12,13}. In \cite{12} author has discussed some aspects
of the violation of the equivalence principle in these theories
for nonlinear $f(R)$.

In this letter we consider the matter dominated universe in the
flat $FRW$ space-time, by using the Palatini formalism of $f(R)$
gravity and by following the so called Noether symmetry approach
\cite{14,15,16}. We look for $f(R)$ cosmological
models which are consistent with Noether symmetry. Metric $f(R)$ cosmology has been examined by this approach \cite{17} but the Palatini formalism hasn't been considered in the
literature yet. It is important to note that for applying the Noether symmetry approach,  we should make a point-like  lagrangian for the cosmological model. In the case of metric formalism, using the method of lagrange multipliers, the Ricci scalar and the cosmic scale factor can be considered as two independent dynamical variables and then by choosing a suitable lagrange multiplier, the cosmological lagrangian takes the point-like form. But in the case of Palatini formalism one cannot use this procedure.  Varying with respect to Ricci scalar leads to an inappropriate lagrange multiplier since the time derivatives of variables appear in the denominator of the lagrange multiplier.  So one cannot construct a canonical effective lagrangian. In this letter, in order to construct a point-like lagrangian,  we use the dynamical equivalence between Palatini $f(R)$ gravity and Brans-Dicke theory.

 The letter is organized as follows.  In Sect. 2, we search for Noether symmetries
 of the action which leads to  explicit forms for $f(R)$. Sect. 3 is
devoted to find out the exact solutions for cosmic scale factor
and the discussion of the various sub-cases.

\section{2.\;\;\;Noether symmetry approach}
Let us begin by introducing the action of the Palatini $f(R)$ theories
\begin{eqnarray}
S=\frac{1}{2k}\int
d^{4}x\sqrt{-g}f(\tilde{R})+S_{m}(g_{\mu\nu},\psi_{m})
\end{eqnarray}
Here $f(\tilde{R})$ is a function of $\tilde{R}=g^{\mu\nu}
R_{\mu\nu}(\tilde{\Gamma})$, where
$\tilde{\Gamma}^{\lambda}_{\mu\nu}$ is the connection. The matter
action $S_{m}$ depends on the matter fields $\psi_{m}$ and
$g_{\mu\nu}$. Varying Eq. (1) with respect to the metric we obtain
\begin{eqnarray}
f'(\tilde{R})\tilde{R}-\frac{1}{2} f(\tilde{R})g_{\mu\nu}= k
T_{\mu\nu}
\end{eqnarray}
where $f'(\tilde{R})=df/d\tilde{R}$. The trace of Eq. (2) is
\begin{eqnarray}
f'(\tilde{R})\tilde{R}-2f(\tilde{R})=k T
\end{eqnarray}
and the variation of Eq. (1) with respect to connection gives
\begin{eqnarray}
\nabla_{\lambda}(\sqrt{-g}f'(\tilde{R})g^{\mu\nu})=0
\end{eqnarray}
so the connection is compatible with
$h_{\mu\nu}=f'(\tilde{R})g_{\mu\nu}$. Therefore we
obtain
\begin{eqnarray}
\tilde{R}=R+\frac{3}{2f'(\tilde{R})}
\partial_{\lambda}f'(\tilde{R})\partial^{\lambda}f'(\tilde{R})-\frac{3}{f'(\tilde{R})}\square f'(\tilde{R})
\end{eqnarray}
where $R$ is Ricci scalar constructed from the Levi-Civita connection
of the metric $g_{\mu\nu}$. One can easily
verify that the action (1) is dynamically equivalent to
\begin{eqnarray}
S=\frac{1}{2k}\int d^{4}x \sqrt{-g}(\Phi R+\frac{3}{2\Phi}
\partial_{\mu}\Phi \partial^{\mu}\Phi -V(\Phi))+S_{m}
\end{eqnarray}
where $\Phi=f'(\tilde{R})$ ,
$V(\Phi)=\chi(\Phi)\Phi-f(\chi(\Phi))$ and $\tilde{R}=\chi(\Phi)$
\cite{18}. This is the action of  Brans-Dicke theory with the
coupling parameter equal to $-\frac{3}{2}$. For matter dominated
cosmology, the matter Lagrangian can be chosen as
$L_{m}=-\rho_{m0}a^{-3}$ where $a$ is  the cosmic scale factor and
$\rho_{mo}$ is a suitable integration constant connected to
matter content.

Using the flat Friedmann-Robertson-Walker metric, the scalar
curvature takes the form
$R=-6(\frac{\ddot{a}}{a}+\frac{\dot{a}^{2}}{a^{2}})$, where the
dot denotes the derivative with respect to time. In order to apply
the Noether symmetry approach, one can easily verify that, in a
$FRW$ manifold, the Lagrangian related to the action (6) takes the
point-like form
\begin{eqnarray}
{\cal L}=12a^{2}\varphi \dot{\varphi}
\dot{a}+6\varphi^{2}\dot{a}^{2}a+6a^{3}\dot{\varphi}^{2}-V(\varphi)a^{3}-2k\rho_{m0}
\end{eqnarray}
in which we have used the redefinition $\Phi\equiv\varphi^{2}$. The equations of motion for $a$
and $\varphi$ are respectively
\begin{eqnarray}
\ddot{\varphi}+\varphi
(\dot{H}+H^{2})+2\dot{\varphi}H+\frac{1}{2}\varphi
H^{2}-\frac{\dot{\varphi}^{2}}{2\varphi}+\frac{V(\varphi)}{4\varphi}=0
\end{eqnarray}
\begin{eqnarray}
\ddot{\varphi}+\varphi (\dot{H}+H^{2})+3\dot{\varphi}H+\varphi
H^{2}=0
\end{eqnarray}
here $H$ is the Hubble parameter. Finally, as a result of general covariance, the energy function associated with the Lagrangian (7) vanishes, that is
\begin{eqnarray}
E_{{\cal L}}=2\dot{\varphi}H+\varphi
H^{2}+\frac{\dot{\varphi}^{2}}{\varphi}+\frac{V(\varphi)}{6\varphi}+\frac{2k\rho_{m0}}{\varphi
a^{3}}=0
\end{eqnarray}

Now, let us introduce the \textit{lift vector field X} \cite{19}
as an infinitesimal generator of the Noether symmetry in the
tangent space $TQ \{a,\dot{a},\varphi,\dot{\varphi}\}$ related to
the configuration space $Q\equiv\{a,\varphi\}$ as follows
\begin{eqnarray}
X=A\frac{\partial}{\partial a}+B\frac{\partial}{\partial
\varphi}+\dot{A}\frac{\partial}{\partial
\dot{a}}+\dot{B}\frac{\partial}{\partial \dot{\varphi}}
\end{eqnarray}
where $A$ and $B$ are unknown functions of the variables $a$ and
$\varphi$. The existence condition for the symmetry, $L_{X}{\cal
L}=0$, leads to the following system of partial differential equations
\begin{eqnarray}
2\varphi A+aB+\varphi^{2}\frac{\partial
A}{\partial\varphi}+a\varphi\frac{\partial A}{\partial
a}+a^{2}\frac{\partial B}{\partial a}+a\varphi\frac{\partial
B}{\partial \varphi}=0
\end{eqnarray}
\begin{eqnarray}
\varphi A+2aB+2a\varphi\frac{\partial A}{\partial
a}+2a^{2}\frac{\partial B}{\partial a}=0
\end{eqnarray}
\begin{eqnarray}
3A+2\varphi\frac{\partial A}{\partial\varphi}+2a\frac{\partial
B}{\partial\varphi}=0
\end{eqnarray}
\begin{eqnarray}
3a^{2}V(\varphi)A+B\frac{dV}{d\varphi}a^{3}=0
\end{eqnarray}
From Eq. (15) we have
\begin{eqnarray}
A=[-\frac{V'(\varphi)}{3V(\varphi)}]Ba
\end{eqnarray}
Substituting (16) into (13), we find that $A=f(\varphi)a^{n}$ and
$-V'/3V=\frac{-2n}{1+2n} \varphi^{-1}$, where $n$ is an
arbitrary number. By substituting these results in (14) we obtain
\begin{eqnarray}
f(\varphi)=\beta \varphi^{n-1}
\end{eqnarray}
where $\beta$ is a constant.These results satisfy Eq. (12) for any arbitrary n. From Eqs. (16)
and (17) we have
\begin{eqnarray}
A=\beta a^{n}\varphi^{n-1}, B=-\frac{(2n+1)\beta}{2n}a^{n-1}\varphi^{n}
\end{eqnarray}
\begin{eqnarray}
V(\varphi)=\lambda
\varphi^{\frac{6n}{1+2n}}=\lambda\Phi^{\frac{3n}{1+2n}}
\end{eqnarray}
where $\lambda$ is a constant. In conclusion, the Noether symmetry
for the Lagrangian (7) exists and the vector field $X$ is
determined by (18) and (11). We can rewrite Eq. (19) as follows
\begin{eqnarray}
\tilde{R}f'(\tilde{R})-f(\tilde{R})=\lambda[f'(\tilde{R})]^{\frac{3n}{1+2n}}
\end{eqnarray}
and we can solve this equation for finding the form of
$f(\tilde{R})$. There exist two series of solutions for this
equation
\begin{eqnarray}
f(\tilde{R})=g(n)\tilde{R}^{\frac{3n}{n-1}}
\end{eqnarray}
\begin{eqnarray}
f(\tilde{R})=\alpha \tilde{R}-\lambda \alpha^{\frac{3n}{1+2n}}
\end{eqnarray}
where
$g(n)=[27n^{3}(\frac{\lambda}{2n+1})^{\frac{2n+1}{n}}]^{\frac{n}{1-n}}(n-1)$.
Solution (22) represents the Hilbert-Einstein action with cosmological
constant. For this action metric and Palatini formalisms coincide.
The existence of Noether symmetry means that there exists a
constant of motion. The constant of motion for Lagrangian (6),
$\Sigma$ , is
\begin{eqnarray}
\Sigma=A \frac{\partial {\cal L}}{\partial \dot{a}}+B
\frac{\partial {\cal L}}{\partial
\dot{\varphi}}=-\frac{6\beta}{n}a^{n+1}\varphi^{n}\frac{d(\varphi
a)}{dt}
\end{eqnarray}
\section{3.\;\;solutions}
\section{3.1\;\; Case\;1\;\;$f(\tilde{R})=g(n)\tilde{R}^{\frac{3n}{n-1}}$}
From Eq. (3) we have
\begin{eqnarray}
\tilde{R}f'(\tilde{R})-2f(\tilde{R})=-k\rho_{m0}a^{-3}
\end{eqnarray}
This equation can be regarded as an equation for $\tilde{R}$ in
terms of the cosmic scale factor, that is
\begin{eqnarray}
\tilde{R}=G(n)a^{\frac{1-n}{n}}
\end{eqnarray}
where $
G(n)=[\frac{k\rho_{m0}(1-n)}{g(n)(n+2)}]^{\frac{n-1}{3n}}$. In
this case the Noether constant is
\begin{eqnarray}
\Sigma=-\frac{3\beta}{2n^{2}}\gamma(n)^{\frac{n+1}{2}}(\dot{a}
a^{\frac{-(n+1)}{2n}})
\end{eqnarray}
in which:
\begin{eqnarray}
\gamma(n)=\frac{3g(n)n}{n-1}G(n)^{\frac{2n+1}{n-1}}
\end{eqnarray}
we can use Eq. (26) to find out the time dependence of the cosmic
scale factor, that is
\begin{eqnarray}
a(t)=(a_{0}+\sigma(n) t)^{\frac{2n}{n-1}}
\end{eqnarray}
where:
\begin{eqnarray}
\sigma(n)=\frac{-n^{2}\Sigma}{3\beta}\gamma(n)^{\frac{-(n+1)}{2}}
\end{eqnarray}
Now we want to consider that whether $a(t)$  obtained
from Noether symmetry, satisfies the field equation of Palatini
$f(\tilde{R})$ cosmology or not. The modified Friedmann equation
in Palatini $f(\tilde{R})$ is (\cite{20})
\begin{eqnarray}
(H+\frac{\dot{f'}}{2f'})^{2}=\frac{1}{6}(\frac{k(\rho+3p)+f}{f'})
\end{eqnarray}
where $\rho$ and $p$ are the energy density and the pressure of
the cosmic fluid respectively. Now by using Eqs. (25) and (30),
after putting $p=0$, we have
\begin{eqnarray}
\dot{a}^{2} a^{\frac{-(1+n)}{n}}=\eta(n)
\end{eqnarray}
where:
\begin{eqnarray}
\eta(n)=\frac{-2n^{2}k
\rho_{m0}}{3\gamma(n)}+\frac{2G(n)n(n-1)}{9}
\end{eqnarray}
so $a(t)$(Eq. (28)) satisfies Eq. (31) if
$\eta(n)=4\sigma(n)^{2}$. This equation can be considered as a
constraint for $\lambda$, $\beta$, $k$ and $n$, so $n$
can be an arbitrary value. Thus for any $n$ there exist Noether
symmetry and one can easily, by using the constraint equation ,
show that the Noether constant of motion is related to the
gravitational constant and matter content of the universe, that is
\begin{eqnarray}
\Sigma^{2} \sim
(c[k\rho_{m0}]^{\frac{2n^{2}+6n+1}{3n}}+d[k\rho_{m0}]^{\frac{2(n+2)}{3}})
\end{eqnarray}
where $c$ and $d$ are functions of $n$.

\section{3.2\;\; Case\;2\;\;$f(\tilde{R})=\alpha \tilde{R}-\lambda \alpha^{\frac{3n}{1+2n}}$}
If $\alpha=1$, then the action is the Hilbert-Einstein action with
cosmological constant $\Lambda=\lambda/2$. It is important to note
that in this case, metric and Palatini formalisms coincide. By
using Eq. (24) we have
\begin{eqnarray}
R=2\lambda+\frac{k\rho_{m0}}{a^{3}}
\end{eqnarray}
and the constant of motion is
\begin{eqnarray}
\Sigma=-\frac{6\beta}{n}(\dot{a}a^{n+1})
\end{eqnarray}
so the scale factor is
\begin{eqnarray}
a(t)=(a_{0}+\sigma(n)(n+2) t)^{\frac{1}{n+2}}
\end{eqnarray}
in which $\sigma(n)=\frac{-n\Sigma}{6\beta}$. For $n=-2$ we have
$a(t)=a_{0}e^{\delta t} $, where $\delta=\frac{\Sigma}{3\beta}$.
The modified Friedmann equation in this case is
$\dot{a}^{2}=\frac{\lambda}{6}a^{2}+\frac{k\rho_{m0}}{3a}$.
 The solution for a(t) which is obtained from Noether symmetry, Eq. (36), satisfies
 the generalized Friedmann equation only if $n=-1/2$, $\lambda=0$ and $\sigma^{2}=\frac{3}{4}k\rho_{m0}$. So in this case
 $a(t)=(a_{0}+\frac{\Sigma}{8\beta} t)^{2/3}$, which is the
 special case of Eq. (28) when $n=-1/2$. Also it is easy to verify that the Noether constant is
 related to gravitational constant and matter density of the
 universe, that is $\Sigma^{2} \sim k\rho_{m0}$.
\section{4.\;\;Concluding remarks}
In this letter , we have considered the Palatini $f(\tilde{R})$
cosmology by a general method. The approach is based on the seek
for Noether symmetry which allows to fix the form of
$f(\tilde{R})$ in a physically motivated manner. In order to
construct a point-like Lagrangian, we have used the dynamical
equivalence between $f(\tilde{R})$ gravity and Brans-Dicke theory, since in the framework of Palatini
$f(\tilde{R})$ gravity, one cannot fix the form of $f(\tilde{R})$ by using the Lagrange multipliers method \cite{17}.

We have shown that this symmetry always exist for $f(\tilde{R})\sim \tilde{R}^{n}$ and the
Noether constant is a function of the Newton's gravitational
constant and the current matter content of the universe. It is interesting to note that in the metric formalism,  the Noether symmetry exists for some limited forms of $f(R)$ such as $R+2\Lambda$, $R+\alpha R^{2}$ and $R^{3/2}$[16]  in the vaccum case.  We know that in vacuum,  Palatini $f(R)$ gravity is equivalent to $GR$ with cosmological constant, so for any arbitrary function of $R$, the Noether symmetry exists. On the other hand, in the matter dominated universe the Noether symmetry exists only for $f(R)\backsim R^{3/2}$  in the metric formalism[17] but in Palatini approach according to Eq. (21), our calculation shows that this symmetry exists considering any arbitrary power of $R$  in the lagrangian.

 A further point which has to be stressed is that Noether symmetry approach
don't allow $f(R)$ forms such as $f(R)=R+\epsilon \phi(R)$, which
have attracted many interest in the literature as the suitable
forms \cite{20,21}. However, this is a satisfactory result in the
sense of the results that have been obtained in \cite{12}, which
imply that any $f(R)$ in Palatini formalism, except linear one,
can violates the equivalence principle. It is also important to
mention that one cannot follow this approach in the metric-affine
formalism of $f(R)$ gravity, since there is no equivalence with
Brans-Dicke theories when the matter action depends on the
connection \cite{18}.\\

\textbf{Acknowledgments}\ \ \    This work is partly supported by a grant from University of Tehran and partly by a grant from center of excellence of department of physics in the structure of matter.

\end{document}